\documentclass{PoS}

\usepackage{graphicx}

\title{Decays $h\to \gamma \gamma, \, \gamma Z$  in the Two Higgs Doublet Model Type III }

\ShortTitle{$h\to \gamma \gamma, \, \gamma Z$ decays in 2HDM-III }

\author{ A. Cordero-Cid \\
      Fac. de Cs. de la
Electr\'onica, Benem\'erita Universidad Aut\'onoma de Puebla, Apdo. Postal 542, Puebla, Pue. 72570, M\'exico. \\
Email: \email{acordero@ece.buap.mx}}

\author{\speaker{J. Hern\'andez-S\'anchez}\thanks{Supported by SNI-CONACYT and PROMEP-SEP grants.}\\
       Fac. de Cs. de la
Electr\'onica, Benem\'erita Universidad Aut\'onoma de Puebla, Apdo. Postal 542, Puebla, Pue. 72570, M\'exico.\\
 Dual C-P Institute of High Energy Physics, Puebla, Pue., M\'exico.\\
        E-mail: \email{jaime.hernandez@correo.buap.mx}}

\author{C. G. Honorato \\
Centro Universitario Valle de Chalco, Universidad Aut\'onoma del Estado de M\'exico, Hermenegildo Galeana No. 3 Col. Maria Isabel, Valle de Chalco. C.P. 56615 Edo. de M\'exico \\
Email: \email{carlos\_honorato@ymail.com}}

\author{S. Moretti \\
School of Physics and Astronomy, University of Southampton, Highfield, Southampton SO17 1BJ, United Kingdom and Particle Physics Department, Rutherford Appleton Laboratory, Chilton, Didcot, Oxon OX11 0QX, United Kingdom \\
E-mail: \email{s.moretti@soton.ac.uk}}

\abstract{We study the enhancement of the branching ratios of the decays $h  \to \gamma \gamma, \, \gamma Z$ in the Two Higgs Doublet Model Type III, assuming a four-zero Yukawa Texture and a general Higgs potential. We show that these processes are very sensitive to the flavor pattern of the Yukawa texture and the structure of the triple coupling $h H^\pm H^\mp$ from  the Higgs potential. We can accomodate the parameters of the model such that one can obtain the $h \to\gamma \gamma$ rates reported by the LHC and at the same time we can  get a $ h \to \gamma Z$ fraction larger than in the SM and within experimental reach. The possibility of obtaining a light charged Higgs boson within the ensuing parameter space and compatible with current experimental measurements is also presented. }

\FullConference{Prospects for Charged Higgs Discovery at Colliders - CHARGED 2014,\\
		16-18 September 2014\\
		Uppsala University, Sweden}

\begin{document}

\section{Introduction}
The flavor sector could also be interesting, as such extensions of the SM could be testable therein. Amongst the latter, tests have been carried out in the most general  version of a 2-Higgs Doublet Model (2HDM) with a Yukawa texture of four-zero, which  can avoid the main flavor physics constraints \cite{HernandezSanchez:2012eg, Felix-Beltran:2013tra},
because this texture form  is the mechanism to control
Flavor Changing Neutral Currents (FCNCs), by taking the Yukawa couplings  proportional to the geometric mean of two fermions masses, $g_{ij} \propto \sqrt{m_i m_j} \chi_{ij}$ \cite{HernandezSanchez:2011ti, DiazCruz:2009ek}. Assuming this setup, which refers to the 2HDM type III (hereafter, 2HDM-III for short), we show that substantial enhancements of the two decay channels  $h\to \gamma \gamma, \, \gamma Z$ are possible. 
\section{The Higgs-Yukawa sector of the 2HDM-III}
2HDMs have two Higgs  doublets of hypercharge $+1$: $\Phi_i^\dag=(\phi _i^-, \, \phi_i^{0*}) $ ($i=1, \, 2$). The most general $SU(2) \times U(1)_Y$ invariant Higgs potential is given by
\begin{eqnarray}
V(\Phi_1, \Phi_2) &=& \mu_1^2 \Phi_1^\dag \Phi_1 +\mu_2^2 \Phi_2^\dag \Phi_2-\bigg( \mu_{12}^2 \Phi_1^\dag \Phi_2 + H. c.\bigg)+\frac{1}{2}\lambda_1 (\Phi_1^\dag \Phi_1)^2 \nonumber \\
&+&
\frac{1}{2}\lambda_2 (\Phi_2^\dag \Phi_2)^2 
+ \lambda_3 (\Phi_1^\dag \Phi_1) (\Phi_2^\dag \Phi_2)
+\lambda_4 (\Phi_1^\dag \Phi_2) (\Phi_2^\dag \Phi_1) \\
&+& \bigg(\frac{1}{2}\lambda_5 (\Phi_1^\dag \Phi_2)^2+
\bigg(  \lambda_6 (\Phi_1^\dag \Phi_1)+\lambda_7 (\Phi_2^\dag \Phi_2)\bigg)  (\Phi_1^\dag \Phi_2)  + H. c. \bigg), \nonumber 
\end{eqnarray}
where all parameters are assumed to be real, including the scalar field  vacuum expectation values $\langle \Phi_i \rangle^\dag =(0, \,v_i)$ ($i=1, \, 2$). Thus,  neither explicit nor spontaneous CP-violation can happen. Besides, when we use a specific four-zero texture for the Yukawa matrices, it plays the role of a flavor symmetry \cite{HernandezSanchez:2011ti, DiazCruz:2009ek}. Therefore, the terms proportional  to $\lambda_6$ and $\lambda_7$ must be preserved. The latter parameters are relevant in one-loop processes through the self-interaction of Higgs bosons \cite{Cordero-Cid:2013sxa,HernandezSanchez:2011fq}.  At the same time, it was shown that radiative contributions to the EW  $\rho = m_W^2/m_Z^2 \cos^2\theta_W$ parameter are large when $(m_{H^\pm} - m_A)$ or $(m_{H^\pm} - m_H)$ are sizable, wherein the parameters $\lambda_{6,7}$ are irrelevant. In fact, the strongest constraints for the most general Higgs potential of 2HDMs come from tree-level unitarity \cite{Ginzburg:2005dt} and one will find that, for $\tan \beta \leq 10$, $|\lambda_{6,7}|\leq 1$, which will be used in this work.
The interactions of the type Higgs-fermion-fermion are derived from the Yukawa Lagrangian
\begin{eqnarray}
{\cal{L}}_Y = - \bigg( Y^{u}_1 \bar{Q}_L {\tilde\Phi_1} u_R+Y^{u}_2 \bar{Q}_L {\tilde\Phi_2} u_R + Y^{d}_1 \bar{Q}_L {\Phi_1} d_R + Y^{d}_2 \bar{Q}_L {\Phi_2} d_R + Y^{l}_1 \bar{L}_L {\Phi_1} \ell_R + Y^{l}_2 \bar{L}_L {\Phi_2} \ell_R \bigg), 
\label{lag-f}
\end{eqnarray}
where ${\tilde\Phi_{1,2}}= i \sigma \Phi_{1,2}^*$. After spontaneous EW Symmetry Breaking (EWSB), we can obtain the fermion mass matrices from eq. (\ref{lag-f}), 
\begin{eqnarray}
M_f= 1/\sqrt{2} (v_1 Y_1^f +v_2 Y_2^f),  \,\,\, \,\,\, f= u, \, d, \, \ell,  \label{Mf}
\end{eqnarray}
where the Yukawa matrices $Y_{1,2}^f$ have the four-zero texture form and are Hermitian. 
Following this definition,
we can compare the Higgs-fermion-fermion ($\phi ff$) coupling of the 2HDM-III with some more standard  2HDMs, e.g. with the 2HDM-II, through the re-difinitions:
\begin{eqnarray}
Y_1^d= \frac{\sqrt{2}}{v \cos \beta} M_d -\tan \beta Y_2^d,
\,\,\,\,\,\,\,\,\,\,\,\,\,\,\,\,
 Y_2^u= \frac{\sqrt{2}}{v \sin \beta} M_u -\cot \beta Y_1^d,
\end{eqnarray}
Obtaining the coupling $\phi ff$ for the 2HDM-III as: $g_{2HDM-III}^{ff \phi} = g_{2HDM-II}^{ff \phi} + \delta g^{ff \phi}$, where $\delta g$ contain the contribution of Yukawa texture. Following this idea we can re-define and get the couplings of $\phi ff$ for  any version of 2HDM plus the contribution of Yukawa texture, as is presented in  \cite{HernandezSanchez:2012eg}:
\begin{eqnarray}
g_{2HDM-III}^{ff \phi} &=& g_{2HDM-any}^{ff \phi} + \delta g^{ff \phi}.
\label{g2hdm}
\end{eqnarray}
When the diagonalization of the Yukawa sector is carried out, the rotated matrices $\tilde{Y}_{1,2}^f$ are
\begin{eqnarray}
\left[\tilde{Y}_{1,2}^f  \right]_{ij}= \frac{\sqrt{m_i^f m_j^f}}{v}  \left[\tilde{\chi}_{1,2}^f  \right]_{ij},
\label{Yij}
\end{eqnarray}
where the $\chi$'s are dimensionless parameters of the model, which have been constrained by several processes of flavor physics such that: $\chi_{kk} \sim O(1)$ and $|\chi_{ij}| \leq 10 ^{-1}, \,\,  (i\neq j)$
\cite{HernandezSanchez:2012eg, Felix-Beltran:2013tra}. Using eqs. (\ref{lag-f}) and (\ref{Yij}) one can get the generic interactions of $H^\pm f_u f_d$ and $h f f$
\begin{eqnarray}
{\cal{L}}_Y &=& - \bigg\{  \frac{\sqrt{2}}{v} \bar{u}_i (m_{d_j} X_{ij} P_R + m_{u_i} Y_{ij} P_L ) d_j H^+  + H.c.\bigg\} 
-\frac{1}{v} \bigg\{   \bar{f}_i m_{f_j} h_{ij} f_j h  \bigg\},
\label{hff}
\end{eqnarray}
where $h_{ij}$,  and $X_{ij}$, $Y_{ij}$  are given by
\begin{eqnarray}
X_{ij} &=& \sum_{l=1}^{3} (V_{CKM})_{il} \bigg \{ X \frac{m_{d_{l}}}{m_{d_{j}}} \delta_{lj} - \frac{f(X)}{\sqrt{2}} \sqrt{ \frac{m_{d_{l}}}{m_{d_{j}}}} \chi_{lj}^{d} \bigg \},
 \,\,\,\,\,\,\,\,\,\,
Y_{ij} = \sum_{l=1}^{3}  \bigg \{ Y  \delta_{il} - \frac{f(Y)}{\sqrt{2}} \sqrt{ \frac{m_{u_{l}}}{m_{u_{i}}}} \chi_{il}^{u} \bigg \} (V_{CKM})_{lj}, \nonumber \\ 
h_{ij}^{d} & =&   \varepsilon_{h}^{d} \delta_{ij} + \frac{ (\varepsilon_{H}^{d} - X \varepsilon_{h}^{d})}{\sqrt{2} f(X)} \sqrt{ \frac{m_{d_{j}}}{m_{d_{i}}}} \chi_{ij}^{d}
\label{hijd}, \,\,\,\,\,\,\,\,\,\,\,\,\,\,\,\,\,\,\,\,
 \,\,\,\,\,\,\,\,\,\,\,\,\,\,\,\,\,\,\,\,
h_{ij}^{u} =  \varepsilon_{h}^{u} \delta_{ij} - \frac{ (\varepsilon_{H}^{u} - Y \varepsilon_{h}^{u})}{\sqrt{2} f(Y)} \sqrt{ \frac{m_{u_{j}}}{m_{u_{i}}}} \chi_{ij}^{u}
\label{XYh}
\end{eqnarray}
with $f(x)= \sqrt{1-x^2}$, the parameters $X$, $Y$ are real and can be related to $\tan \beta$ or $\cot \beta$ and $\varepsilon_{h}^{u}$ with the mixing angles $\alpha$ and $\beta$, as is shown in the Table I.  
This Lagrangian could also represent a Multi-Higgs Doublet Model (MHDM) or an Aligned 2HDM (A2HDM) with an additional flavor symmetry, as also suggested in \cite{HernandezSanchez:2012eg}.
{\small
\begin{table}
\begin{tabular}{|c|c|c|c|c|c|c|c|c|c|}
\hline 
2HDM-III &$ X$ & $Y$ & $Z$ & $\varepsilon_{h}^{u}$ & $\varepsilon_{h}^{d}$ & $\varepsilon_{l}^{d}$ & $\varepsilon_{H}^{u}$ & $\varepsilon_{H}^{d}$ & $\varepsilon_{H}^{l}$ \\ 
\hline 
2HDM-I-like & $- \cot \beta$ & $\cot \beta$  & $-\cot \beta$ & $c_{\alpha}/ s_{\beta}$ &$ c_{\alpha}/ s_{\beta}$ &  $c_{\alpha}/ s_{\beta}$ & $s_{\alpha} / s_{\beta}$ & $s_{\alpha} / s_{\beta}$  & $s_{\alpha} / s_{\beta}$  \\ 
\hline 
2HDM-II-like & $\tan \beta$  & $\cot \beta$ & $\tan \beta$ & $c_{\alpha}/ s_{\beta}$ & $-s_{\alpha} / c_{\beta}$  & $-s_{\alpha} / c_{\beta}$  & $s_{\alpha} / s_{\beta}$  & $c_{\alpha}/ c_{\beta}$ & $c_{\alpha}/ c_{\beta}$\\ 
\hline 
2HDM-X-like & $- \cot \beta$ & $\cot \beta$ & $\tan \beta$ & $c_{\alpha}/ s_{\beta}$ & $c_{\alpha}/ s_{\beta}$ & $-s_{\alpha}/ c_{\beta}$ & $s_{\alpha}/ s_{\beta}$ & $s_{\alpha}/ s_{\beta}$ &$ c_{\alpha}/ c_{\beta}$ \\ 
\hline 
2HDM-Y-like &  $\tan \beta$ & $\cot \beta$ & $ -\cot \beta$ &$ c_{\alpha}/ s_{\beta} $& $-s_{\alpha}/ s_{\beta}$ & $c_{\alpha}/ s_{\beta}$ & $s_{\alpha}/ s_{\beta}$ & $ c_{\alpha}/ c_{\beta}$ &$ s_{\alpha}/ s_{\beta}$ \\ 
\hline 
\end{tabular} 
\caption{Parameters $\varepsilon_{h}^{f}$, X and Y defined in the Yukawa interactions for four versions of the 2HDM-III. Here $s_\alpha = \sin\alpha$, $c_\alpha = \cos\alpha$, $s_\beta = \sin\beta$, $c_\beta = \cos\beta$}
\end{table} }

\section{Decays $h\to \gamma \gamma, \, \gamma Z$ in the 2HDM-III }
Once we have isolated the surviving parameter space for this version of 2HDM (type III)  \cite{Cordero-Cid:2013sxa,HernandezSanchez:2012eg}, we study the Branching Ratios (BRs) of the two channels $h\to \gamma \gamma$
and $  \gamma Z$.  We consider particular encarnations of our 2HDM-III (2HDM-I, 2HDM-II, 2HDM-X, 2HDM-Y), by 
suitably taking into account new contributions given by the flavor physics represented through the four-zero texture of the Yukawa matrices. As the experimental results suggest a scenario close to SM-like Higgs signal, we choose $\beta - \alpha \sim \pi/2$ and $\mu_{12} \sim v$, $m_A=200$ GeV and $m_H=250$ GeV. Moreover, we show that the greatest enhancement of the BRs is when $\lambda_{6}= -\lambda_{7}$ \cite{Cordero-Cid:2013sxa}. Finally, we analyze the so-called $R$ parameters:
\begin{eqnarray}
R_{\gamma X} = \frac{\sigma(gg \to h)_{\rm 2HDM-III} \times {\rm BR}(h \to \gamma X)_{\rm 2HDM-III}}{\sigma(gg \to h)_{\rm SM} \times {\rm BR}(h \to \gamma X)_{\rm SM}} \, \, \, (X= \gamma, Z).
\end{eqnarray}
In the case of a fermiophobic $h$ state, the $gg\to h$ production mode must be replaced by either vector boson fusion or Higgs-strahlung, for which
 $R = {\rm BR}(h \to \gamma X)_{\rm 2HDM-III}/ {\rm BR}(h \to \gamma X)_{\rm SM}$.

\begin{figure}
\centering
\includegraphics[width=3in]{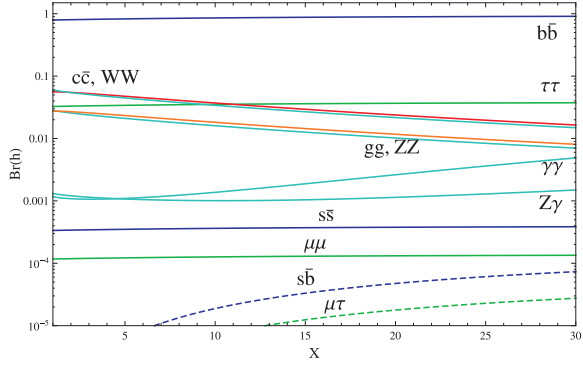} 
\caption{BRs vs $X$, of all decay channels of the light CP-even Higgs boson, for the 2HDM-III like-II. The parameters used are: $\chi_{kk}^f=1$ ($K=1, \, 2, \, 3$),  $\chi_{23}^u=-0.5$, $\chi_{23}^d=-0.035$, $m_{H^\pm}=200$ GeV.}
\label{fig1}
\end{figure}

\begin{figure}
\centering
\includegraphics[width=2.8in]{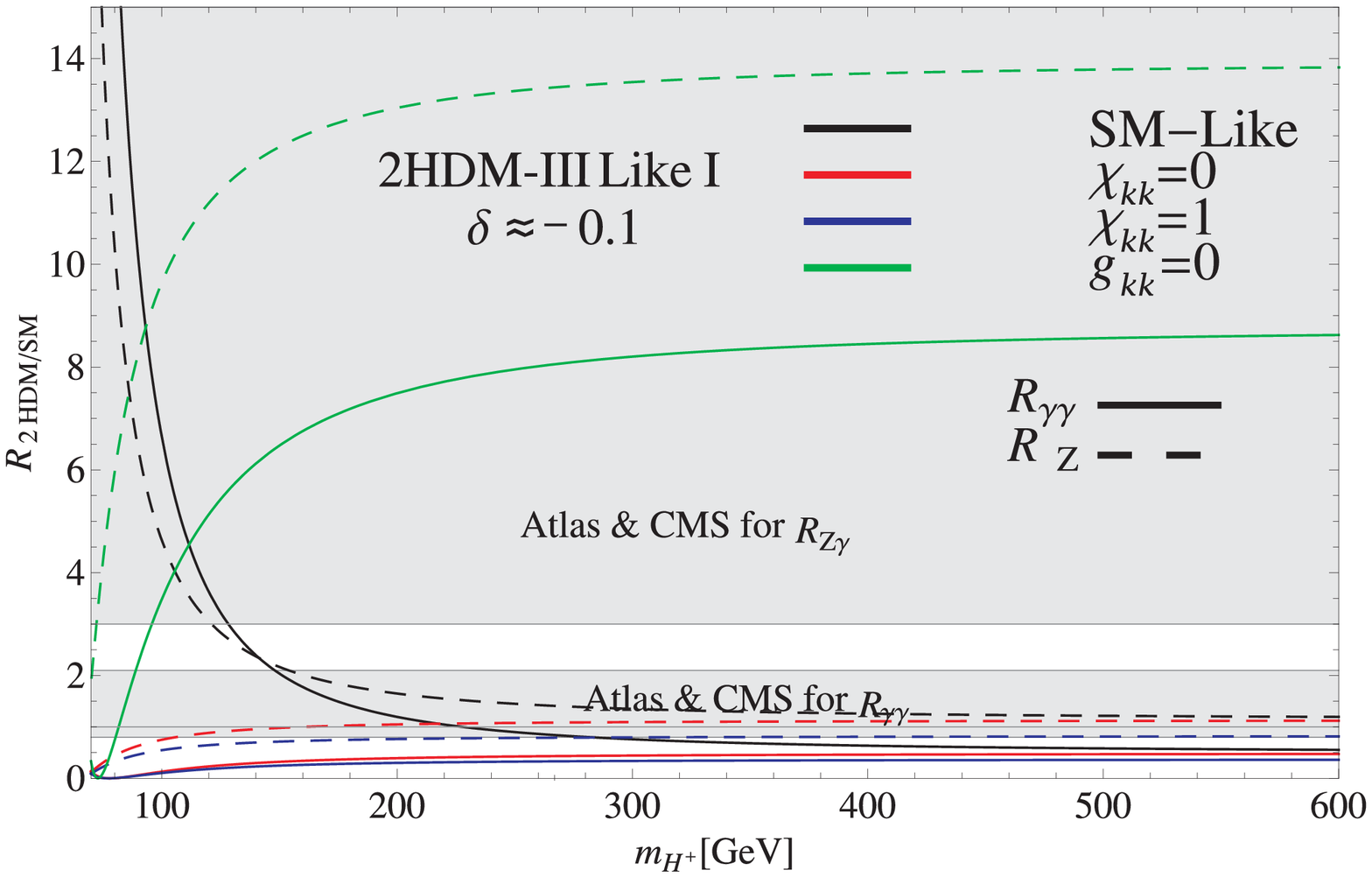} \includegraphics[width=2.8in]{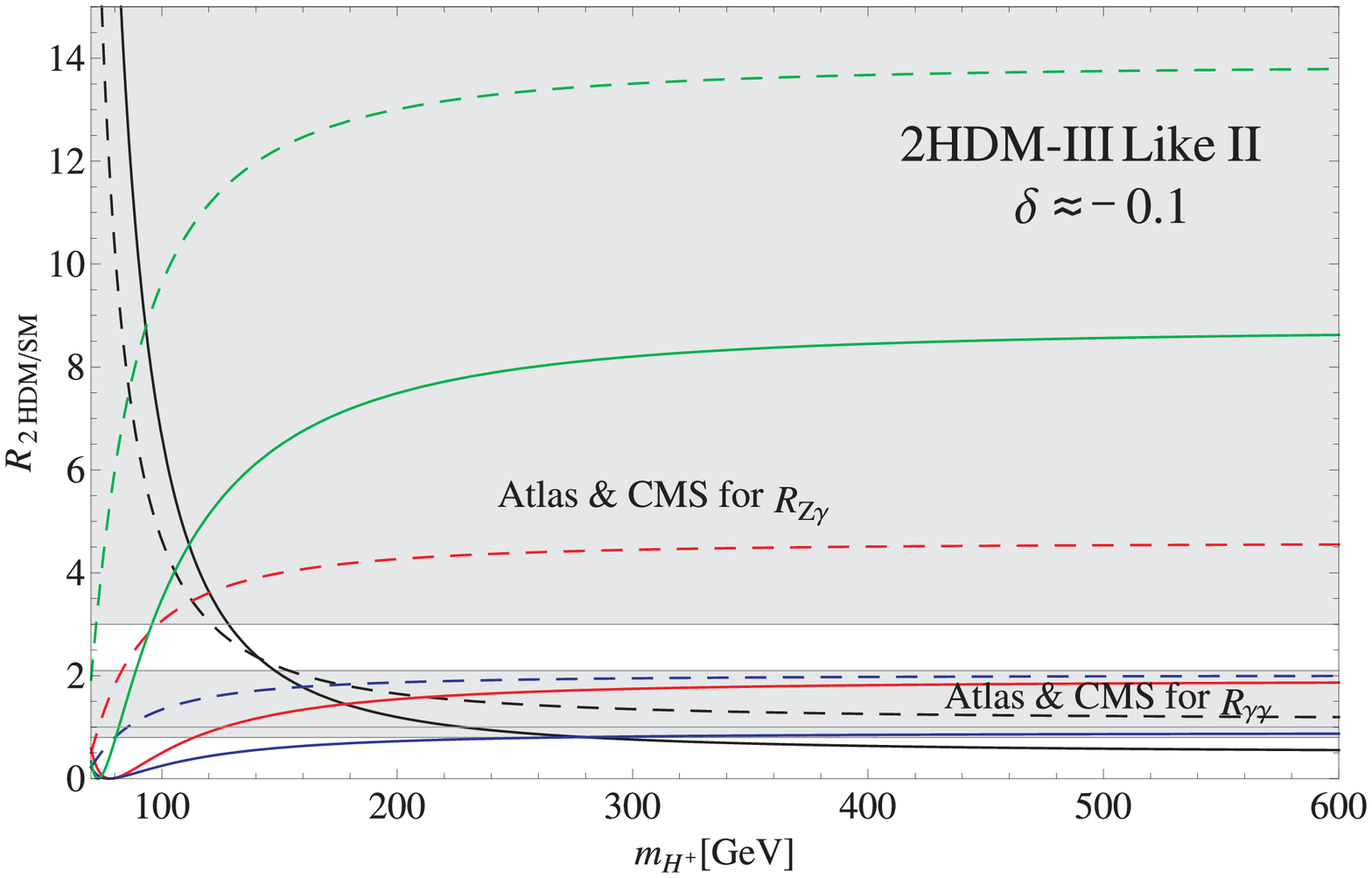} \\
\includegraphics[width=2.8in]{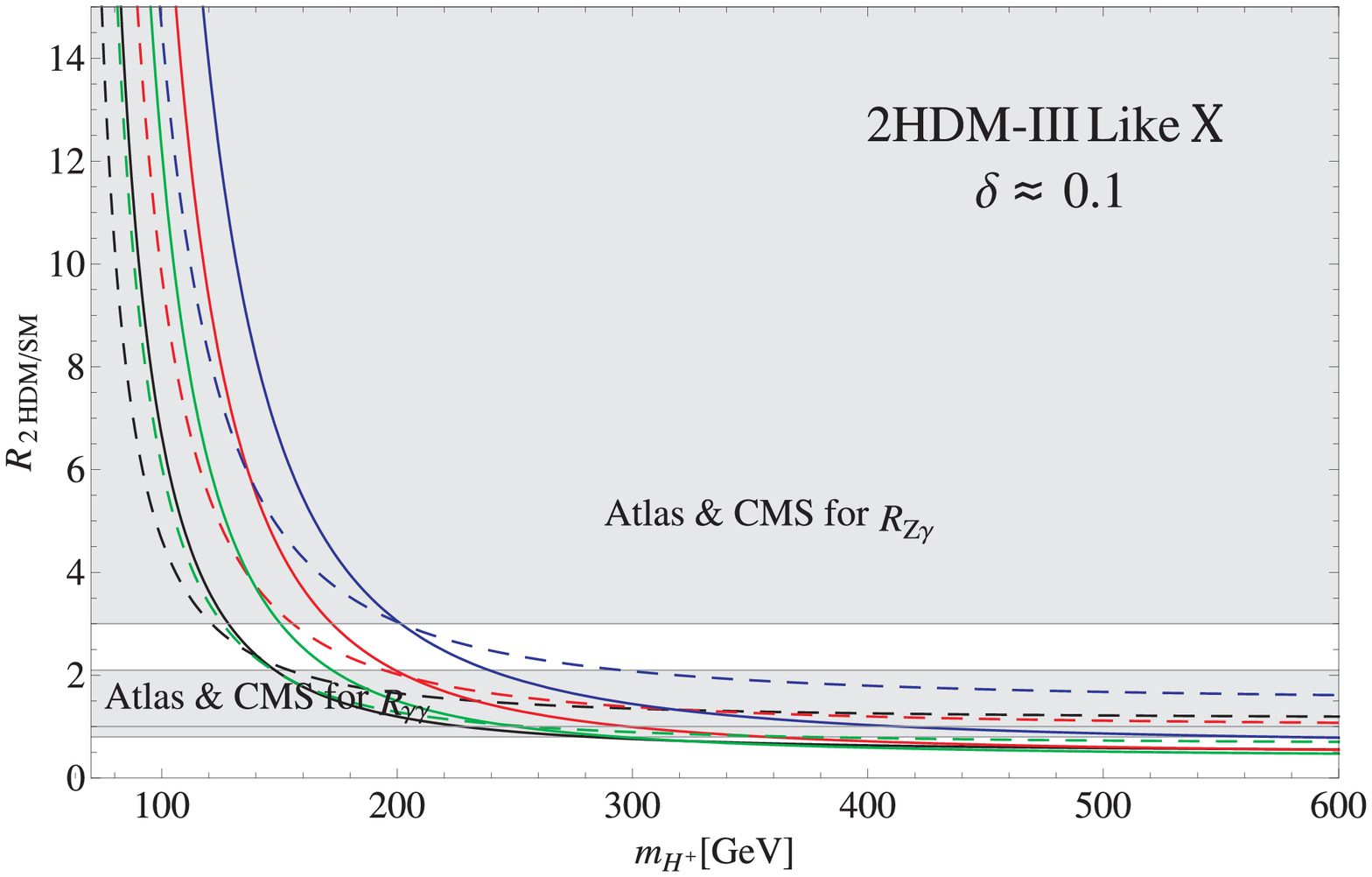} \includegraphics[width=2.8in]{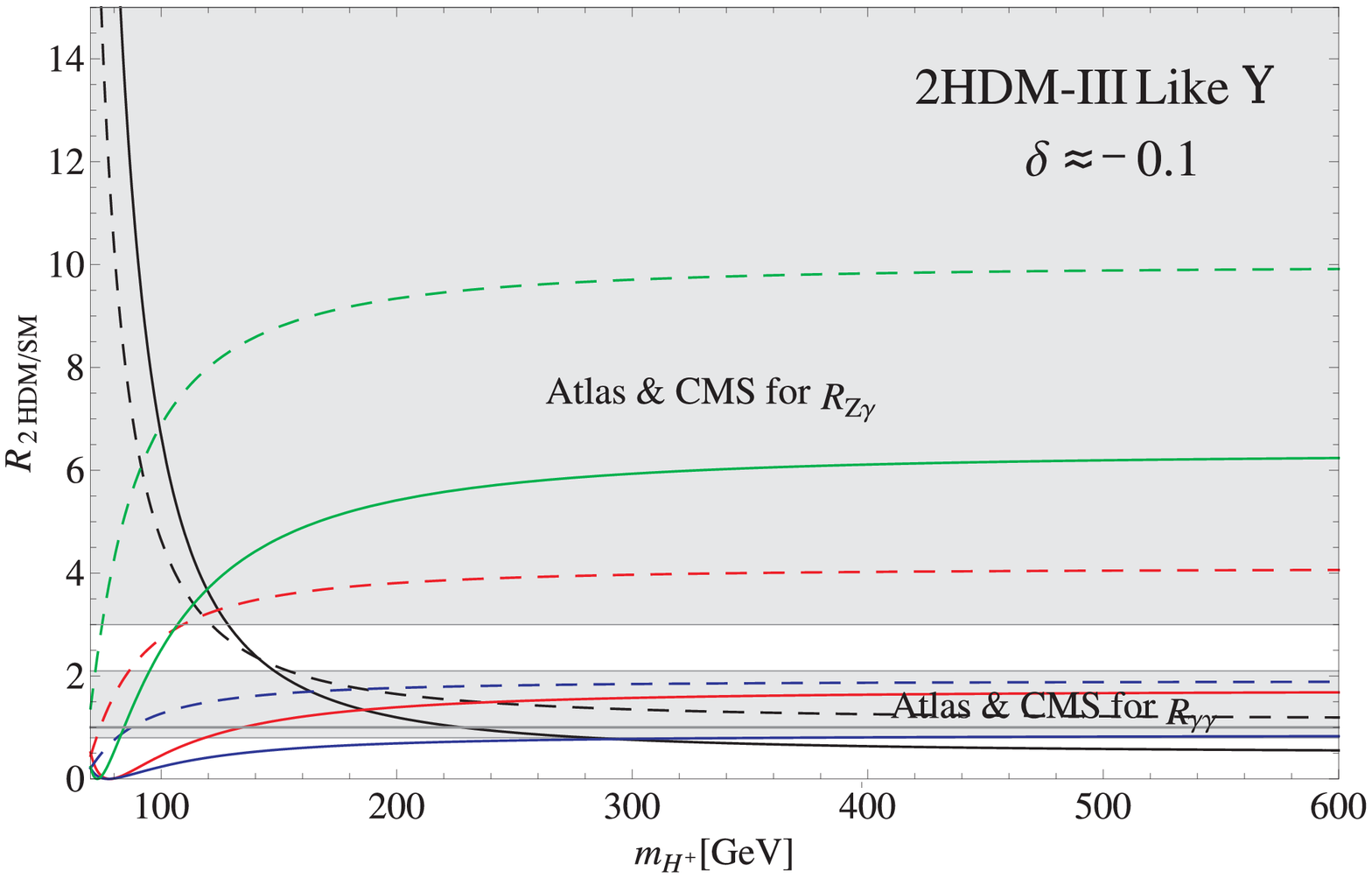} 
\caption{$R_{\gamma \gamma}$ (solid line) and $R_{\gamma Z}$ (dashed line) with respect to the charged Higgs boson mass, where the shaded areas are the fits to the results from the LHC. The other parameters are given in the legends and described in the text. The solid and dashes black  line represent a scenario close to SM like, with $R\sim 1$.}
\label{fig2}
\end{figure}

\begin{figure}
\centering
\includegraphics[width=2.1in]{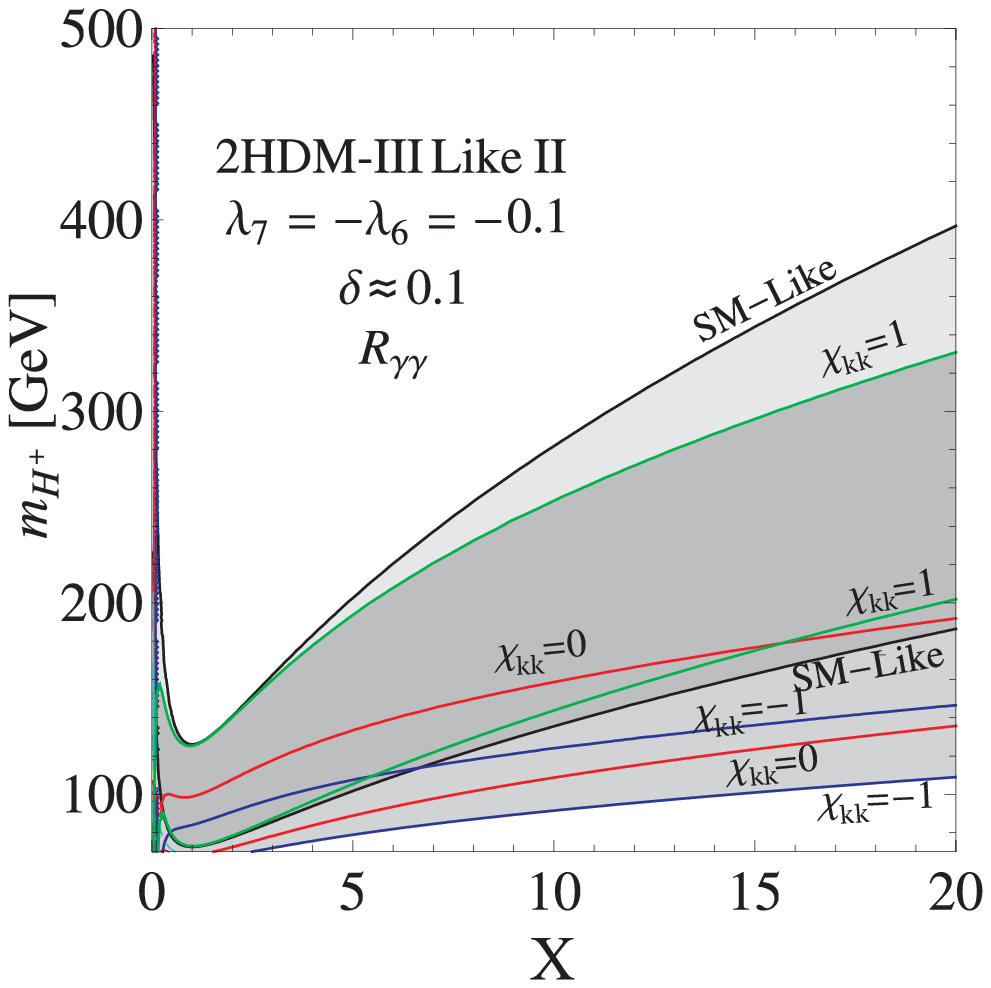} \includegraphics[width=2.1in]{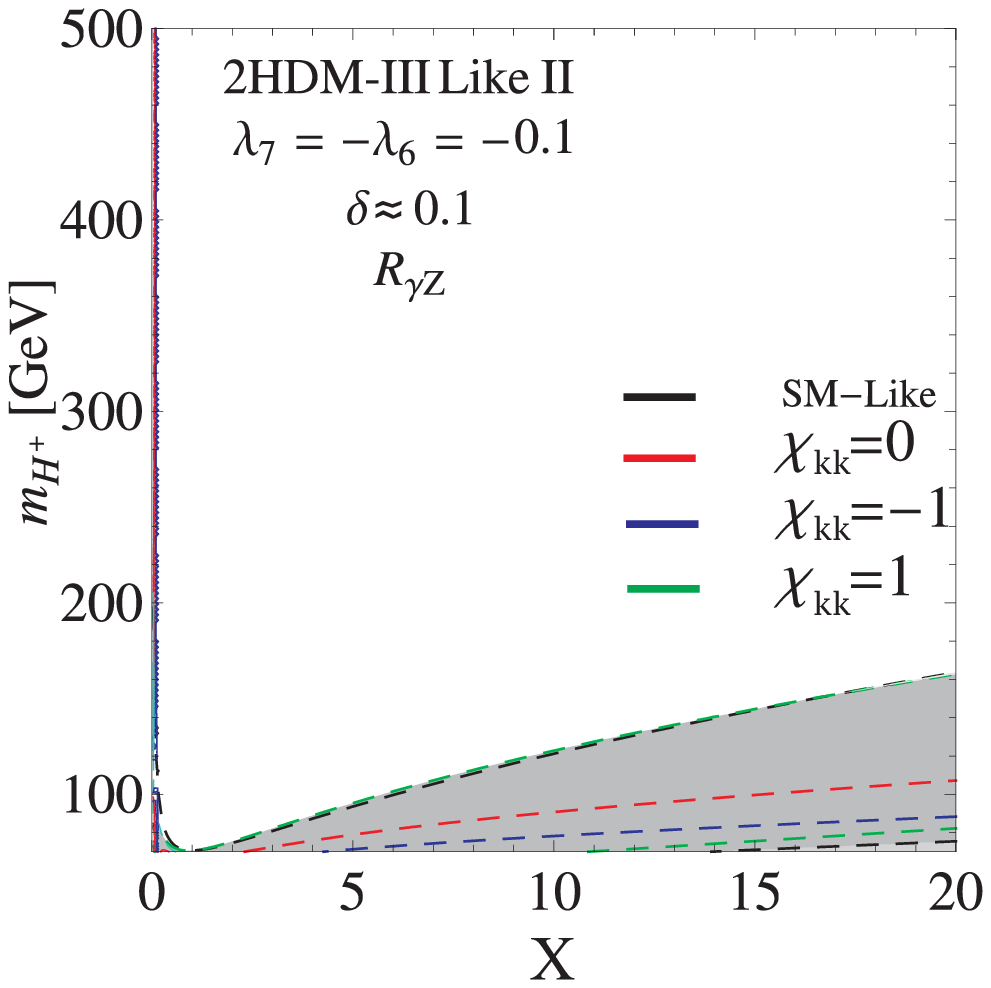} \\
\includegraphics[width=2.1in]{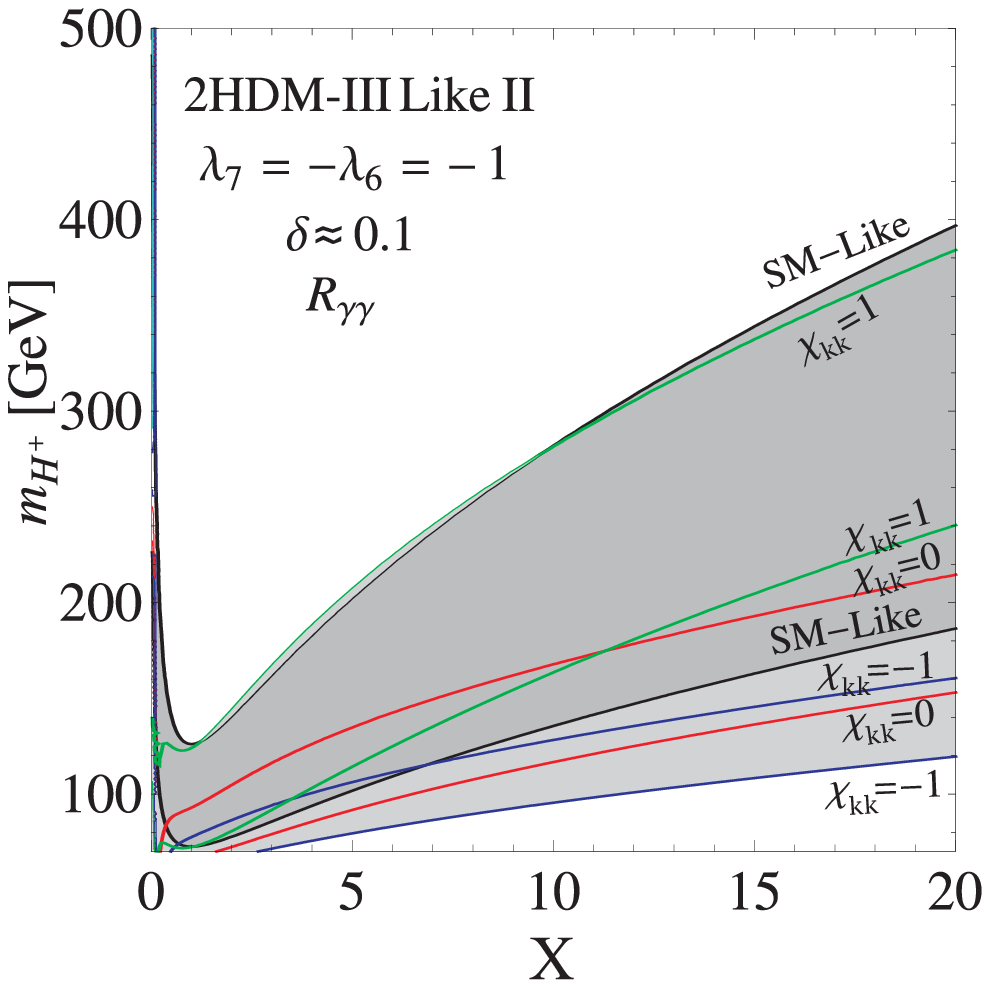} \includegraphics[width=2.1in]{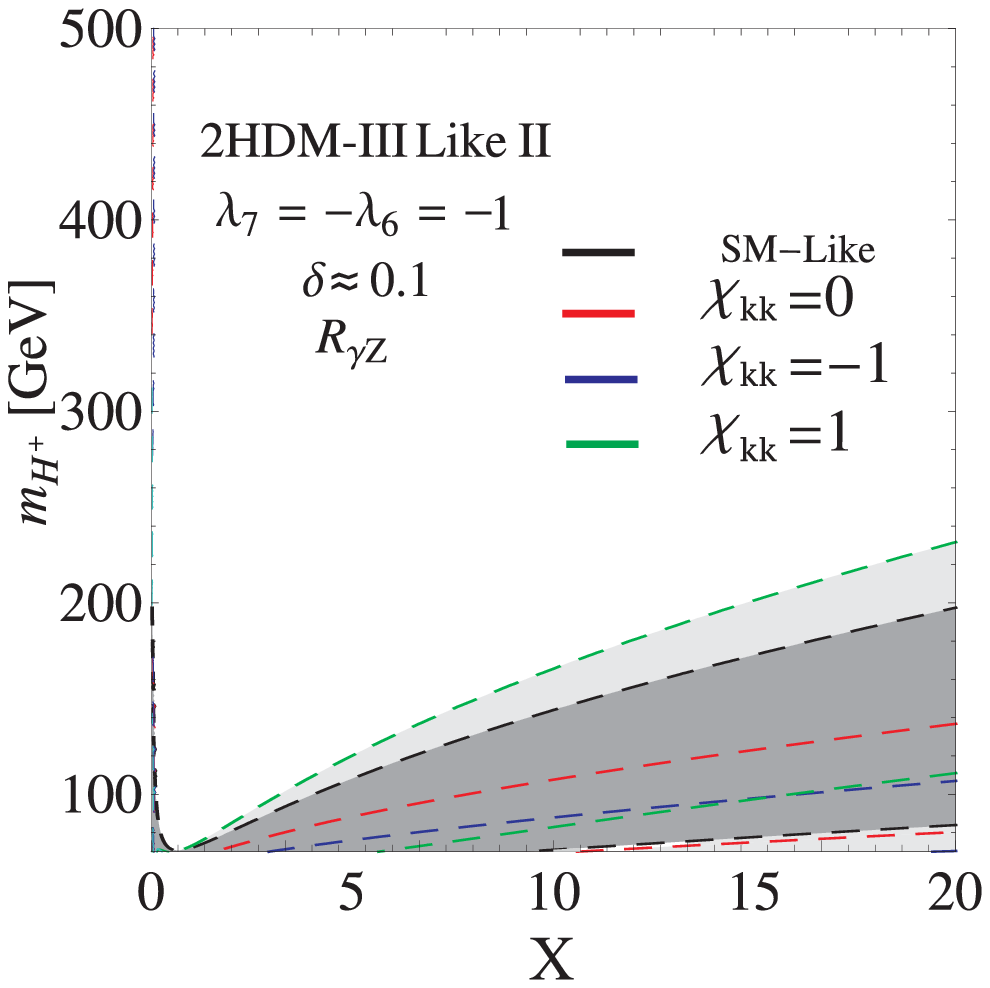} 
\caption{Allowed parameter space in the  $m_{H^\pm} $-$X$ plane for $R_{\gamma \gamma} $ (solid line) and $R_{\gamma Z}$ (dashed line), according to LHC data (the shaded areas enclosed by lines of the same color are the permitted regions). }
\label{fig3}
\end{figure}

Now we discuss the results for the $ h$ decays. Firstly, in Fig.~\ref{fig1} we show that
the BR$(h \to \gamma \gamma)$ is very sensitive to the $X$ parameter, which is introduced in eq. (\ref{hff}),
for the case of the 2HDM-III like-II. For large values of $X$, the BR receives an enhancement of one order of magnitude with respect to the SM rate. However, this behavior is contrary to the experimental results from the LHC. In contrast, for medium values of $X$ ($X< 15$), the BR is under control and is consistent with the measurements from the LHC data (gray region). Besides, over the same parameter space region, the possibility of a light charged Higgs boson is compatible with all such data.
Secondly, in Fig.~\ref{fig2} we present the behavior of $R_{\gamma \gamma}$ (solid line) and $R_{\gamma Z}$ (dashed line)  with respect to the charged Higgs boson mass. The shaded areas (gray region) are the fits to the results from the LHC. We study the following scenarios: the black line is  a  SM-like $h$ state ($\beta -\alpha = \pi/2$) and $R\sim 1 $ in the approximate decoupling limit \footnote{One can see in the Fig. \ref{fig2} that in the approximate decoupling limit R is around 1 and is not  1 exactly, because we only are running the charged Higgs mass. However when all masses and parameters of the model are decoupled, we obtain the decoupling limit exactly 1.};  the red line represents the case where the Yukawa couplings receive contributions from the  $\chi$ parameters ($\chi_{kk}^f=0$, $\chi_{23}^u=-0.5$, $\chi_{23}^d=-0.35$); finally, the green line shows the fermiophobic case. Here, we look at the following relevant scenarios: 2HDM-III like-II and -Y with $\chi_{kk}^f=0$, $\chi_{23}^u=-0.5$, $\chi_{23}^d=-0.35$ and the fermiophobic scenario for the 2HDM-III like-I, -II and -Y.  For all cases we allow for a charged Higgs boson that can be very light, $m_{H^\pm} \geq 110$ GeV. We find a decay rate for $h\to \gamma Z$ mode which is an order of magnitude larger than the one obtained in the SM.
Thirdly, we show in Fig.~\ref{fig3} the allowed $m_{H^\pm} $-$X$ plane for two values of $\lambda_{6,7}$. We present the two relevant signatures for the scenario 2HDM-III like-II. In the plots we contrast these to  the permitted  parameter space, after taking account the LHC data. One can see that the final state $\gamma Z$ is  most constrained than the $\gamma \gamma $ one. In particular, both modes demand $ m_{H^\pm} \leq 160$ GeV ($ m_{H^\pm} \leq 230$ GeV) for $X =20$  and $\lambda_6=-\lambda_7=0.1$ ($\lambda_6=-\lambda_7=1$). Actually, for $X$ around 15, one can find  that $ m_{H^\pm} < m_t$, so that the charged Higgs state is copiously produced in top quark decays. Thus, as we have intimated previously, a light charged Higgs boson could be a manisfestation of the 2HDM-III \cite{Cordero-Cid:2013sxa, HernandezSanchez:2012eg}.
\section{Conclusions}
We have shown some benchmarks where the 2HDM-III parameters space considered is consistent with all current experimental constraints. We find that, assuming a four-zero Yukawa texture, one can have a light charged Higgs boson accessible at the LHC and that the decay channels $h\to \gamma \gamma$ and $ \gamma Z$ are very sensitive to the flavor structure represented by the Yukawa texture and a general Higgs potential.


\begin{thebibliography}{99}
\bibitem{Aad:2012tfa} 
  G.~Aad {\it et al.}  [ATLAS Collaboration],
  Phys.\ Lett.\ B {\bf 716}, 1 (2012)
  [arXiv:1207.7214 [hep-ex]].

\bibitem{Chatrchyan:2012ufa} 
  S.~Chatrchyan {\it et al.}  [CMS Collaboration],
  Phys.\ Lett.\ B {\bf 716}, 30 (2012)
  [arXiv:1207.7235 [hep-ex]].

\bibitem{ATLAS:2013mma} 
  [ATLAS Collaboration],
  ATLAS-CONF-2013-014, ATLAS-COM-CONF-2013-025.

\bibitem{CMS:yva} 
  [CMS Collaboration],
  CMS-PAS-HIG-13-005.

\bibitem{Ferreira:2011aa} 
  P.~M.~Ferreira, R.~Santos, M.~Sher and J.~P.~Silva,
  Phys.\ Rev.\ D {\bf 85}, 077703 (2012)
  [arXiv:1112.3277 [hep-ph]].


\bibitem{Akeroyd:2012ms} 
  A.~G.~Akeroyd and S.~Moretti,
  Phys.\ Rev.\ D {\bf 86}, 035015 (2012)
  [arXiv:1206.0535 [hep-ph]].

\bibitem{Akeroyd:2012rg} 
  A.~G.~Akeroyd and S.~Moretti,
  PoS CHARGED {\bf 2012}, 035 (2012)
  [arXiv:1210.6882 [hep-ph]].

\bibitem{Cordero-Cid:2013sxa} 
  A.~Cordero-Cid, J.~Hernandez-Sanchez, C.~G.~Honorato, S.~Moretti, M.~A.~Perez and A.~Rosado,
  JHEP {\bf 1407}, 057 (2014)
  [arXiv:1312.5614 [hep-ph]].

\bibitem{HernandezSanchez:2012eg} 
  J.~Hernandez-Sanchez, S.~Moretti, R.~Noriega-Papaqui and A.~Rosado,
  JHEP {\bf 1307}, 044 (2013)
  [arXiv:1212.6818].

\bibitem{Felix-Beltran:2013tra} 
  O.~F\'elix-Beltr\'an, F.~Gonz\'alez-Canales, J.~Hern\'andez-S\'anchez, S.~Moretti, R.~Noriega-Papaqui and A.~Rosado,
  Phys.\ Lett.\ B {\bf 742}, 347 (2015)
  [arXiv:1311.5210 [hep-ph]].

\bibitem{HernandezSanchez:2011ti} 
  J.~Hernandez-Sanchez, L.~Lopez-Lozano, R.~Noriega-Papaqui and A.~Rosado,
  Phys.\ Rev.\ D {\bf 85}, 071301 (2012)
  [arXiv:1106.5035 [hep-ph]].

\bibitem{DiazCruz:2009ek} 
  J.~L.~Diaz-Cruz, J.~Hernandez--Sanchez, S.~Moretti, R.~Noriega-Papaqui and A.~Rosado,
  Phys.\ Rev.\ D {\bf 79}, 095025 (2009)
  [arXiv:0902.4490 [hep-ph]].

\bibitem{HernandezSanchez:2011fq} 
  J.~Hernandez-Sanchez, C.~G.~Honorato, M.~A.~Perez and J.~J.~Toscano,
  Phys.\ Rev.\ D {\bf 85}, 015020 (2012)
  [arXiv:1108.4074 [hep-ph]].

\bibitem{Ginzburg:2005dt} 
  I.~F.~Ginzburg and I.~P.~Ivanov,
  Phys.\ Rev.\ D {\bf 72}, 115010 (2005)
  [hep-ph/0508020].

\end{thebibliography}
\end{document}